\documentclass[conference,10pt,letterpaper]{IEEEtran}
\IEEEoverridecommandlockouts

\usepackage[normalem]{ulem}
\usepackage[english]{babel}
\usepackage{amsmath,amsfonts,amssymb}
\usepackage{algorithm,algorithmic}
\usepackage{xurl}
\usepackage{graphicx}
\usepackage{xcolor}
\usepackage{subfigure}
\usepackage{booktabs}
\usepackage{multirow}
\usepackage{hyperref}
\usepackage{enumitem}
\usepackage{esvect}
\usepackage{mathbbol}
\definecolor{backgroundbox}{rgb}{0,0,0}
\def\BibTeX{{\rm B\kern-.05em{\sc i\kern-.025em b}\kern-.08em
    T\kern-.1667em\lower.7ex\hbox{E}\kern-.125emX}}

\newcommand{\HH}{\textcolor{black}}

\begin{document}
\title{QueueTrack: Framework for Detecting the Onset of Queueing in Mobile Networks}
\author{\IEEEauthorblockN{Hamid Hassani, Anudeep Karnam, Kishor Joshi, Sonia M. Heemstra, Ignas G. Niemegeers, George Exarchakos}
\IEEEauthorblockA{Eindhoven University of Technology, The Netherlands\\
{\small \tt \{h.hassani,s.s.a.karnam,k.c.joshi,sheemstradegroot,i.g.m.niemegeers,g.exarchakos\}@tue.nl}
}
}
\maketitle
%
\begin{abstract}
In mobile networks, timely and accurate detection of the onset of queueing at path bottlenecks remains challenging for end-to-end congestion control algorithms (CCAs), where end hosts have limited visibility into queue status. Traditional congestion signals, such as packet loss, round-trip time (RTT), and one-way delay (OWD), have well-known limitations for early queue buildup detection. This paper presents QueueTrack, a practical end-host framework to detect the onset of queueing in mobile networks. 
QueueTrack computes a queue buildup indicator, denoted QBI, from endpoint-observable packet-level timing statistics, without requiring tight synchronization between end hosts or modifications to the existing network infrastructure. We develop and assess the complete QueueTrack machine learning (ML) pipeline, including data collection, labeling, model training, and testing, on a 5G testbed built on the OpenAirInterface platform, under a range of scenarios with competing TCP traffic. 
The experimental results show that QueueTrack can reliably detect the onset of queueing at the 5G base station using low-complexity classification models such as logistic regression. 
For example, QueueTrack achieves an F1 score of approximately $80\%$ in a challenging mixed TCP/UDP scenario. 
These findings confirm that QueueTrack can serve as a practical framework for developing CCAs that target low delay, as well as for network telemetry and network monitoring in mobile networks. 
\end{abstract}
\section{Introduction} \label{sec:intro}
Emerging time-critical applications such as extended reality, teleoperation, and cloud gaming demand both high data rate and low latency communication. 
For such immersive services, data rate may need to range from $100$Gbps to $1$Tbps, with latency below $1$ms~\cite{Saad20}. However, these requirements cannot be fully met by current 5G networks~\cite{Saad20,Akyildiz22,She23}, and many network capabilities are expected to grow substantially over the next decade to support emerging data-intensive applications~\cite{Adeia23,SK23}. 

While 3GPP and IEEE 802.11 standardization activities have primarily focused on physical and link layers, recent studies highlight the importance of transport-layer protocols, particularly congestion control algorithms (CCAs), in enabling real-time applications~\cite{Hassani21,Xie22}. 
The need for major upgrades to current CCAs is also reflected in initiatives such as Google BBR~\cite{BBR} and L4S~\cite{RFC9330}.

The choice of an appropriate congestion signal\footnote{An indicator used by CCAs to detect congestion within the network.} is a key challenge to CCA design, since the signal should detect the onset of queueing at the path bottleneck before excessive queueing delay for effective rate adaptation. 
Existing CCAs primarily rely on packet loss or delay measurements as congestion signals. 
However, these signals may prove insufficient for next-generation applications that require both high data rate and low latency~\cite{Vleeschauwer24}. 
Packet loss is not ideal for low delay operation since it typically triggers rate reduction only after queue overflow has occurred, leading to high queueing delay (QD)~\cite{Tanenbaum21}. 
Moreover, CCAs that rely on packet loss as the congestion signal, such as TCP Cubic~\cite{Ha08}, are unable to distinguish whether packet loss is caused by queue overflow or by wireless impairments such as degraded channel conditions, interference, or handovers. 

The use of packet delay as a congestion signal in CCAs such as TCP BBR~\cite{BBR} can be inaccurate for inferring queue buildup at the path bottleneck. For instance, round-trip time (RTT) is known to be inaccurate when there is queueing in the reverse path, i.e., asymmetric forward and reverse path conditions~\cite{Chefrour21}. One-way delay (OWD) provides a more direct forward-path measurement, but it is challenging to measure robustly due to, e.g., clock offset and skew between the sender and receiver~\cite{Hassani21}. 

There are also alternative approaches for mobile networks, such as decoding radio-control information using network-wide sniffing~\cite{PBE20,Xie22,Wan24a,Wan24b} to estimate available capacity, or using explicit congestion notification (ECN) protocols \cite{RFC3168} at bottleneck queues. 
However, these strategies are often impractical due to deployment and upgrade difficulties. 
For example, L4S~\cite{RFC9330}, which relies on ECN feedback from the base station (BS), requires extensive updates to network infrastructure and end-user devices, making widespread adoption challenging. 

To address the above challenges, we present \emph{QueueTrack}, a deployable end-host framework that exposes implicit information about queue states using packet timing statistics, without requiring access to the queue status at the BS. 
Instead of relying on packet loss, RTT, or OWD to detect queue buildup at the BS, we propose a novel metric, the queue buildup indicator (\emph{QBI}). 
Using measurements at the user equipment (UE), QBI provides a practical signal for identifying the onset of queueing at the path bottleneck. 
Note that, in a 5G edge architecture, the bottleneck queue typically lies at the radio link control (RLC) layer of the BS~\cite{Irazabal24}.
QueueTrack uses simple machine learning (ML) models, i.e., logistic regression (LR), support vector machine (SVM), and multilayer perceptron (MLP), to avoid unnecessary CPU overhead. 
Furthermore, it does not require any modifications to the existing network infrastructure and can be integrated with low-delay CCAs or used for network telemetry, runtime diagnostics, and network monitoring. 

The main contributions of this paper are as follows:
\begin{itemize}
    \item We introduce QBI, a novel metric for tracking the queue status at the RLC layer of the BS in 5G edge networks where the air interface is the path bottleneck. QBI is computed from packet-level timing statistics observed at the end hosts and does not require tight synchronization between end hosts. 

    \item We develop QueueTrack, a user-space end-host software framework that uses QBI and ML models to detect the onset of queueing at the BS. QueueTrack is flexible and easy to deploy, without requiring changes to the existing network infrastructure.

    \item We conduct an extensive measurement campaign to build a dataset and evaluate QueueTrack on a 5G testbed using the OpenAirInterface (OAI) platform~\cite{OAI} and under a broad range of scenarios, including single- and multi-UE settings, coexistence with widely used CCAs, robustness to uplink feedback loss, and measurement perturbations.
    
    \item The experimental results demonstrate that QueueTrack can accurately detect the onset of queueing under realistic and complex 5G testbed conditions. Moreover, the evaluated learning models, i.e., LR, SVM, and MLP, generalize to previously unseen mixed-flow scenarios with TCP Cubic and BBR cross-traffic, achieving F1 scores of $80\%$.
\end{itemize}

The remainder of the paper is organized as follows. 
In Section~\ref{sec:related}, we review the related work. 
We present the system architecture of QueueTrack in Section~\ref{sec:design}, and in Section~\ref{sec:metric}, we introduce the QBI metric. 
In Section~\ref{sec:eval}, we describe the testbed, measurement process, and dataset construction. 
Next, we present the QueueTrack learning pipeline in Section~\ref{sec:pipeline}. 
In Section~\ref{sec:results}, we discuss the performance evaluation results. 
We conclude the paper in Section \ref{sec:conclusion}. 
\section{Related Work} \label{sec:related}
The primary objective of CCAs is to estimate the available bandwidth and adjust the send rate, e.g., by updating the TCP congestion window (CWND) size. 
To this end, the literature has proposed several congestion signals, which are generally divided into loss- and delay-based approaches. 
Loss-based CCAs, such as TCP BIC~\cite{Xu04} and TCP Cubic~\cite{Ha08}, use packet loss as the congestion signal, whereas delay-based CCAs, such as TCP BBR~\cite{BBR} and TCP Copa~\cite{Arun18}, rely on delay measurements, e.g., RTT, to infer queue buildup at the path bottleneck.

ECN protocols, such as L4S~\cite{RFC9330}, use queue dynamics, e.g., queueing delay, as an explicit congestion signal. While such mechanisms can provide timely congestion feedback, they require changes in existing network equipment. This often implies widespread software upgrades across network nodes, such as access points and BSs. This process is notably expensive and resource-intensive, which can ultimately restrict the agile integration of innovative solutions. 

There are alternative approaches that rely on network-wide sniffing to measure wireless channel bandwidth, as outlined in~\cite{PBE20,Xie22} for LTE and~\cite{Wan24a,Wan24b} for 5G. 
They deploy passive sniffers in different locations to capture and decode the control channel over the air interface. 
However, decoding control channels is not a practical solution~\cite{piStream15}, and such approaches may suffer from performance limitations under realistic network conditions. 

\cite{claw17} presents a tool called CLAW that can estimate the LTE cell-wide resource utilization based on the physical layer information, such as reference signal received quality (RSRQ), available on unmodified commodity (but rooted) mobile handsets. CLAW needs around one RTT to estimate the available bandwidth. 
However, results in~\cite{claw17} show that the predictive power of signal strength is relatively limited~\cite{PBE20}. Furthermore, energy measurements have typically resulted in significant estimation errors due to interference~\cite{Xie22}.

ExLL~\cite{ExLL18} analyzes the patterns in packet reception in short time windows, i.e., at an LTE subframe granularity, to infer the LTE link bandwidth. This solution relies on end-to-end statistics, so it is sensitive to network dynamics and may inaccurately estimate the available bandwidth~\cite{PBE20}. 

Cesar~\cite{Shin25} is another advanced delay-based CCA for cellular networks, designed to maintain low delay and high data rate by reducing the impact of non-congestive delays, such as scheduling and processing delays at the BS. 
While Cesar has been evaluated in commercial LTE and 5G networks, the authors in~\cite{Shin25} focus primarily on performance comparisons with other CCAs and do not report detailed resource-utilization behavior. 
Therefore, as with other delay-based approaches, resource-utilization concerns remain for Cesar. 

Biscay~\cite{Larrea26} is a recently proposed CCA for mobile networks. 
It relies on fine-grained radio indicators extracted from the device chipset to estimate the available bandwidth, and then combines this estimate with RTT to adjust the CWND. 
However, downlink and uplink are often asymmetric due to, e.g., scheduling policies; as a result, RTT is known to be an inaccurate indicator for inferring congestion~\cite{Pathak08,Chefrour21}. 
Moreover, not all chipsets expose such detailed radio information to upper layers, which limits the practicality and deployability of this approach. 

The authors in~\cite{Hassani25a,Hassani25b} introduce physical resource block (PRB) utilization as a practical congestion signal for mobile networks and show that it is closely coupled with queue buildup at the RLC layer of the BS. 
The proposed CCA, i.e., RAPID, achieves significant delay reduction with moderate goodput degradation and negligible packet loss~\cite{Hassani25a}. 
However, RAPID requires abstract information about the number of active network flows, which limits its applicability mainly to private networks where such information is available. 
QueueTrack addresses this limitation by investigating whether queue dynamics can be inferred from end-host packet timing and delivery observations, without requiring direct access to the queue state at the BS. 
\section{System Model} \label{sec:design}
\begin{figure}[tbp]
    \centering
    \includegraphics[width=0.95\linewidth]{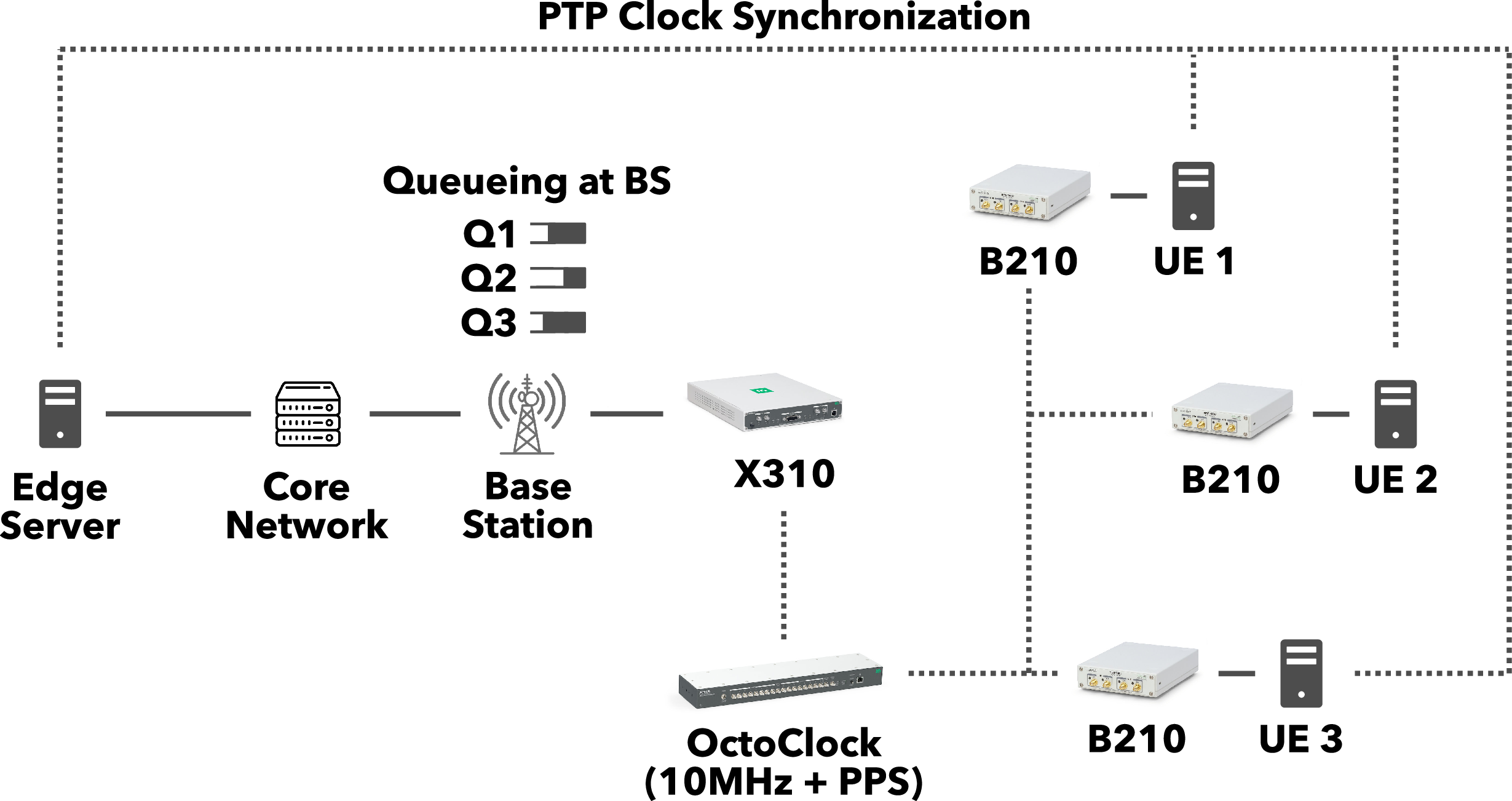}
    \caption{5G edge testbed setup using the OAI platform and Ettus universal software radio peripheral (USRP) devices. The path bottleneck in this edge transport architecture lies at the RLC layer of the BS, as indicated on the schematic. The OctoClock provides time and frequency synchronization for the USRPs.}
    \label{fig:system}
    \vspace{-0.20cm}
\end{figure}

We consider a 5G edge architecture, as illustrated in Fig.~\ref{fig:system}, where the edge server (ES) is collocated with the BS at the network edge. 
The path bottleneck in this setup lies at the RLC layer of the BS. 
As depicted in Fig.~\ref{fig:system}, the BS adheres to a per-UE queueing policy, where each UE has its own dedicated queue at the BS~\cite{Irazabal24}. 
Note that the major source of end-to-end delay is typically the QD at the RLC layer of the BS ~\cite{Flavien22,Irazabal24,RFC9330,Hassani25b}. 

Our interest is to detect the onset of queue buildup at the BS, as depicted in Fig.~\ref{fig:system}, in a practical and accurate manner via an innovative approach deployed in user space \HH{as an application running outside the operating system's kernel.} 
To this end, QueueTrack is implemented at the end hosts and does not require modifications to the 5G standard or network infrastructure. 
To infer the onset of queueing, the ES timestamps each packet by writing its send time in the packet payload using the system kernel time, enabling the UE to retrieve this information upon reception. 
The UE then uses this sender-side timing information together with packet arrival timing statistics to compute QBI and periodically report the resulting measurements back to the ES. 
\HH{QueueTrack can similarly be applied to uplink traffic by swapping the roles of the ES and UE.} 

The perfect synchronization between the ES and the UEs is required only to precisely measure OWD as an evaluation reference for latency behavior and to derive offline ground-truth labels for supervised learning. 
OWD is not used as an input feature for the models or any other QueueTrack component. 
Furthermore, we assume that the RLC layer operates in unacknowledged mode, which is used for real-time services. 
\section{QBI Metric: Definition and Interpretation} \label{sec:metric}
QueueTrack infers queue buildup by comparing, over each feedback interval, the estimated number of packets offered to the BS with the number of packets received at the UE. 
When the offered packet count persistently exceeds the delivered packet count, packets are likely building up in the RLC queue, whereas an excess of delivered packets indicates that an existing RLC queue backlog is likely being depleted. 
\subsection{Packet Arrival and Departure Processes}
At the UE side, time is partitioned into intervals of duration $\delta$ seconds. 
The $j$th feedback interval is defined as $[t_j, t_j + \delta)$, where $t_j$ denotes its start time. 
We let $A_i^{\mathrm{BS}}(t)$ denote the cumulative number of packets offered to the BS for UE $i$ up to time $t$, and $D_i^{\mathrm{UE}}(t)$ the cumulative number of packets delivered to UE $i$ up to time $t$. 
Hence, for the $j$th interval, the number of offered packets is
\begin{equation}
\Delta A_{i,j}^{\mathrm{BS}} = A_i^{\mathrm{BS}}(t_j + \delta) - A_i^{\mathrm{BS}}(t_j)
\end{equation}
whereas the corresponding number of delivered packets is
\begin{equation}
\Delta D_{i,j}^{\mathrm{UE}} = D_i^{\mathrm{UE}}(t_j + \delta) - D_i^{\mathrm{UE}}(t_j)
\end{equation}

Measuring $\Delta A_{i,j}^{\mathrm{BS}}$ would require modifications to the BS or precise timing alignment between the ES, BS, and UE. 
To address this challenge, QueueTrack uses an end-host estimate of $\Delta A_{i,j}^{\mathrm{BS}}$ over the $j$th feedback interval, denoted by $M_{i,j}$:
\begin{equation} \label{eq:m_approx}
\Delta A_{i,j}^{\mathrm{BS}} \approx M_{i,j}
\end{equation}
On the other hand, the delivered packet count, $\Delta D_{i,j}^{\mathrm{UE}}$, is measured directly at the UE and is denoted by $N_{i,j}$:
\begin{equation} \label{eq:n_approx}
\Delta D_{i,j}^{\mathrm{UE}} = N_{i,j}
\end{equation}

Next, we denote by $s_1, s_2, \ldots, s_n$ the send timestamps of packets received by UE $i$ during the $j$th feedback interval. 
The inter-packet departure time (IPDT) is the time gap between consecutive packet departures at the ES. 
For $n > 1$, the empirical mean IPDT for UE $i$ during the $j$th interval, denoted by $\mu_{\mathrm{IPDT}_i}(\delta_j)$, is defined as follows:
\begin{equation}
\mu_{\mathrm{IPDT}_i}(\delta_j) = \frac{1}{n-1} \sum_{k=1}^{n-1}(s_{k+1} - s_k)
\end{equation}

We assume that backhaul is reliable, with negligible packet loss, and that its transport delay is small relative to the duration of the feedback interval $\delta$. 
Furthermore, we assume that the send rate remains constant within each feedback interval. 
Under these assumptions, the number of packets offered to the BS can be approximated as:
\begin{equation}
{M}_{i,j} = \frac{\delta}{\mu_{\mathrm{IPDT}_i}(\delta_j)}
\end{equation}
Note that bursty traffic, jitter, or rapid rate variations within a reporting interval can make $\mu_{\mathrm{IPDT}}(\delta)$ noisy. QueueTrack mitigates this by relying on short-history QBI features instead of single-interval estimates. 
\subsection{Packet-Count Mismatch}
Let $B_{i,j}$ denote the number of packets that have been offered to the BS for UE $i$ but have not yet been delivered to the UE by the end of the $j$th feedback interval. 
This quantity is not the exact RLC queue status, but it captures the packet-count mismatch that drives backlog buildup. Its dynamics can be written as:
\begin{equation} \label{eq:logical_workload_evolution}
B_{i,j} = B_{i,j - 1} + \Delta A_{i,j}^{\mathrm{BS}} - \Delta D_{i,j}^{\mathrm{UE}}
\end{equation}
If packets are discarded due to buffer overflow or retransmission failure, the gap between the number of offered and delivered packets not only captures the increase in $B_{i,j}$, but also accounts for packets that never reach the destination. Here, we focus on scenarios in which such packet losses are negligible. 

Substituting~\eqref{eq:m_approx} and~\eqref{eq:n_approx} into~\eqref{eq:logical_workload_evolution}, we obtain:
\begin{equation} \label{eq:estimated_workload_evolution}
B_{i,j} \approx B_{i,j - 1} + M_{i,j} - N_{i,j}
\end{equation}
Therefore, the packet-count mismatch over the $j$th interval is
\begin{equation} \label{eq:workload_drift}
\begin{aligned}
\Delta B_{i,j} &= B_{i,j} - B_{i,j - 1} \\
& \approx M_{i,j} - N_{i,j}
\end{aligned}
\end{equation}

Thus, $M_{i,j} - N_{i,j}$ reflects queue dynamics rather than the instantaneous queue occupancy. 
If $M_{i,j} > N_{i,j}$, the number of packets offered exceeds the number of packets delivered, implying that packets are building up at the BS. 
This situation can occur when the UE receives a smaller share of radio resources due to cross-traffic, degraded channel conditions, or increased hybrid automatic repeat request (HARQ) retransmissions. 
If $M_{i,j} < N_{i,j}$, more packets are delivered than offered within the interval, indicating that an existing backlog is being drained. 
Finally, if $M_{i,j} \approx N_{i,j}$, the offered and delivered packet counts are roughly balanced, indicating that the newly offered packets neither build a new backlog nor increase the existing backlog at the BS. 
\subsection{Queue Buildup Indicator} \label{subsec:QBI}
While the packet-count mismatch in~\eqref{eq:workload_drift} reflects queue dynamics, its magnitude depends on the offered load. 
Thus, QueueTrack normalizes this mismatch using the estimated number of offered packets. 
For intervals with $M_{i,j} > 0$, QBI is defined as
\begin{equation} \label{eq:qbi_definition}
\mathrm{QBI}_{i,j} = \frac{M_{i,j} - N_{i,j}}{M_{i,j}}
\end{equation}
Note that QBI is upper bounded by $1$ but has no lower bound because $N_{i,j}$ can exceed $M_{i,j}$ during periods when the queue is being drained. 
\begin{figure}[tbp]
    \centering
    \includegraphics[width=0.55\columnwidth]{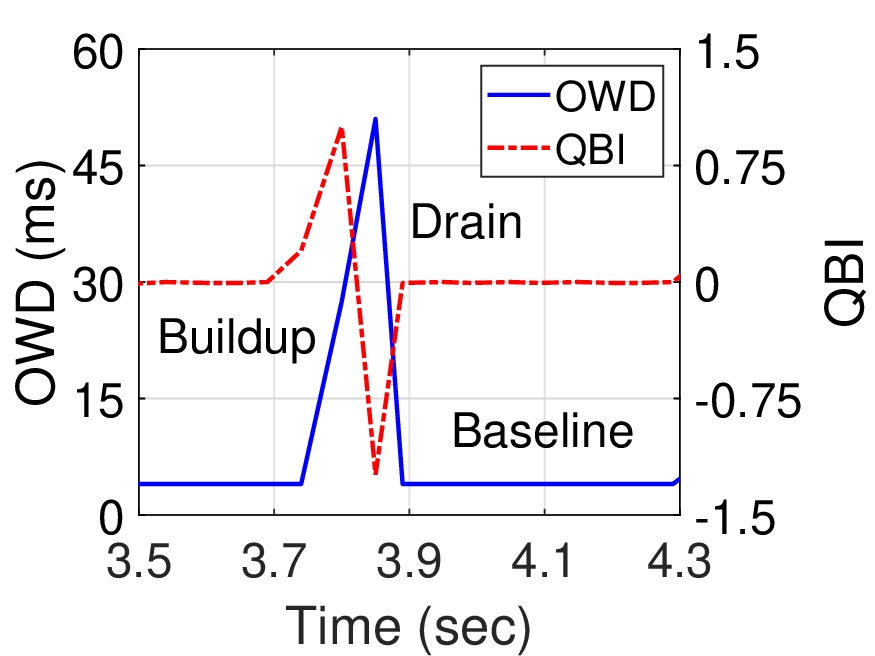}
    \caption{Illustrative evolution of OWD and QBI during queue buildup and depletion. Each sample corresponds to a feedback report with $\delta = 50$ms.} 
    \label{fig:qbi}
    \vspace{-0.20cm}
\end{figure}

Fig.~\ref{fig:qbi} provides an illustrative example of how QBI changes relative to OWD as the queue builds up and drains. 
In the absence of a persistent queue backlog, OWD remains low and QBI stays close to its baseline. 
As packets begin to accumulate, QBI increases, indicating that the offered packet count exceeds the delivered packet count. 
However, when the backlog starts to deplete, QBI decreases and becomes negative before OWD returns to its baseline, showing that more packets are being delivered than newly offered during this interval. 
As the queue backlog clears, both OWD and QBI return to their baseline behavior. 
This behavior shows that QBI is not only correlated with elevated delay, but also reflects the transition dynamics of queue buildup and depletion, \HH{supporting the use of short-history QBI features for detecting the onset of queueing.} 
\section{Experimental Setup and Dataset} \label{sec:eval}
In this section, we describe the testbed, the measurement process, and the dataset construction. 
\subsection{5G Testbed} \label{sec:setup}
The experimental testbed is a 5G standalone network based on the OAI project (branch 2024.w38) for the core, BS, and UEs. It is configured to operate in the n78 frequency band using TDD mode with $6$/$3$/$1$ D/U/M configuration, $40$MHz channel bandwidth with $\text{PRB}_{\text{max},\text{slot}} {=} 106$, single layer transmission, and numerology $1$ ($30$KHz subcarrier spacing with $0.5$ms slot duration). 
This setup adheres to the default configurations of OAI.

We use the \texttt{oai-ext-dn} container of OAI as the ES, which runs on the same machine as the 5G core and the BS. Each UE runs on a dedicated machine. 
The machines have identical hardware specifications (Intel Core i9-14900K CPUs with $128$GB of DDR5 RAM) and run Ubuntu 22.04.5 LTS. The Ethernet NICs are connected through a MikroTik CRS309 switch and synchronized over a fiber-optic LAN via the precision time protocol (PTP) using the \texttt{ptp4l} program\footnote{\texttt{ptp4l} is a Linux implementation of the PTP protocol according to IEEE standard 1588.}~\cite{ptp4l}. 
Moreover, we use the \texttt{phc2sys} program~\cite{phc2sys} to synchronize the system clock to the PTP hardware clock. 

The BS is connected to a USRP X310 software-defined radio (SDR), while the UEs use USRP B210 SDRs to communicate over the air. The USRPs run the open-source UHD driver (v4.6). Furthermore, all USRPs are synchronized using an OctoClock-G CDA-2990 to achieve a high-accuracy time and frequency reference, see Fig. \ref{fig:system}. 
The UEs are placed at various positions, resulting in different channel conditions. Moreover, since the n78 band is the commercial 5G frequency band in the region where the experiments were conducted, there is always potential interference from nearby mobile BSs and UEs. 

We develop the server- and UE-side codes in Python for our prototype implementation of QueueTrack. The ES generates iPerf-like UDP traffic towards the UEs and inserts a sender-side timestamp into the packet payloads. Since the machines are tightly synchronized via the PTP protocol, this timestamp can be used to precisely measure the OWD at the UE. Furthermore, for cross-traffic, we generate iPerf-like TCP flows. 
\subsection{Measurement and Data Collection} \label{sec:measurement}
We collect measurement traces through downlink experiments from the ES to the UE. In the single-UE baseline setting, the ES generates iPerf-like UDP traffic at a fixed send rate for $180$s. 
\HH{We use a fixed packet size such that each IP datagram fits within the path maximum transmission unit (MTU) to avoid fragmentation.} 
The UE sends feedback reports to the ES every $\delta$ seconds. We sweep the UDP send rate over $\{10,20,30,40\}$Mbps and $\delta \in \{25,50,75,100\}$ms, and execute each configuration as an independent run. 
The UE records the send rate, receive rate, OWD, and QBI for each feedback report. 

Furthermore, we conduct experiments with TCP cross-traffic, where the ES generates one or more iPerf-like TCP flows using TCP Cubic and/or TCP BBR while the UDP flow under test remains active. 
The resulting dataset is used to evaluate the generalization of our models under complex and previously unseen congestion scenarios. 
TCP packets are timestamped using the same method as UDP packets, enabling OWD measurement for comparative analysis. 

For the multi-UE setup, the UDP flow under test operates at a fixed send rate in $\{10, 15, 20\}$Mbps, while additional UEs simultaneously generate competing TCP downlink traffic. 
The evaluated scenarios are summarized in Table \ref{tab:multiue}, where bracketed values denote the active time interval of each flow. For instance, $[30, 150]$s indicates that the flow starts at $t = 30$sec and terminates at $t = 150$sec. 
Traces are collected for feedback intervals $\delta \in \{25, 50, 75, 100\}$ms. 
Recall that all multi-UE traces are withheld from training and used exclusively for evaluation. 
\begin{table}[t]
\centering
\caption{Evaluation scenarios with TCP cross-traffic.}
\label{tab:multiue}
\resizebox{\columnwidth}{!}{%
\begin{tabular}{l l l}
\hline
\textbf{Scenario} & \textbf{TCP} & \textbf{UDP} \\
\hline
1 Cubic (1C)          & $[30, 150]$s                            & $[0, 180]$s \\
1 BBR (1B)            & $[30, 150]$s                            & $[0, 180]$s \\
2 Cubic (2C)          & both: $[30, 150]$s                      & $[0, 210]$s \\
1 Cubic + 1 BBR (1C+1B)
                      & Cubic: $[30, 150]$s, BBR: $[60, 180]$s & $[0, 210]$s \\
\hline
\end{tabular}
}
\end{table}
\subsection{Dataset Preparation and Labeling} \label{sec:dataset}
The development of a persistent queue backlog corresponds to congestion and increases QD. 
This leads to an increase in OWD above its baseline value\footnote{The baseline OWD is below $10$ms in our testbed.}, which is primarily determined by processing delays at the BS and UE, the transmission time interval (TTI), and alignment delay~\cite{Hassani25b}. 
We therefore define the onset of queueing as the transition from a low-delay regime to a persistent excess-delay regime, rather than isolated spikes in OWD~\cite{Goyal20}. 

We use a threshold-based labeling approach to derive the ground-truth congestion status from OWD. 
For each observation $k$, the label is defined as $y_k = \mathbb{1}\{\text{OWD}_k > \tau\}$, where $\mathbb{1}\{\cdot\}$ represents the indicator function and $\tau$ denotes the delay threshold. 
Measurements with OWD exceeding $\tau$ are labeled as congested ($1$), while all remaining observations are labeled as non-congested ($0$). 

To evaluate both performance within the training regime and generalization to unseen regimes, we partition the dataset by traffic scenario. 
The training set contains only single-flow UDP traces. 
For each training trace, samples are ordered chronologically and then split: the first $75\%$ is used for model training, and the remaining $25\%$ serves as a temporal holdout set drawn from the same regime. 
The multi-flow traces, i.e., those with TCP Cubic and/or BBR cross-traffic, are excluded from training and used exclusively for final testing to evaluate whether QueueTrack generalizes to unseen traffic dynamics. 
\section{QueueTrack Learning Pipeline} \label{sec:pipeline}
\begin{figure}[tbp]
    \centering
    \includegraphics[width=0.35\textwidth]{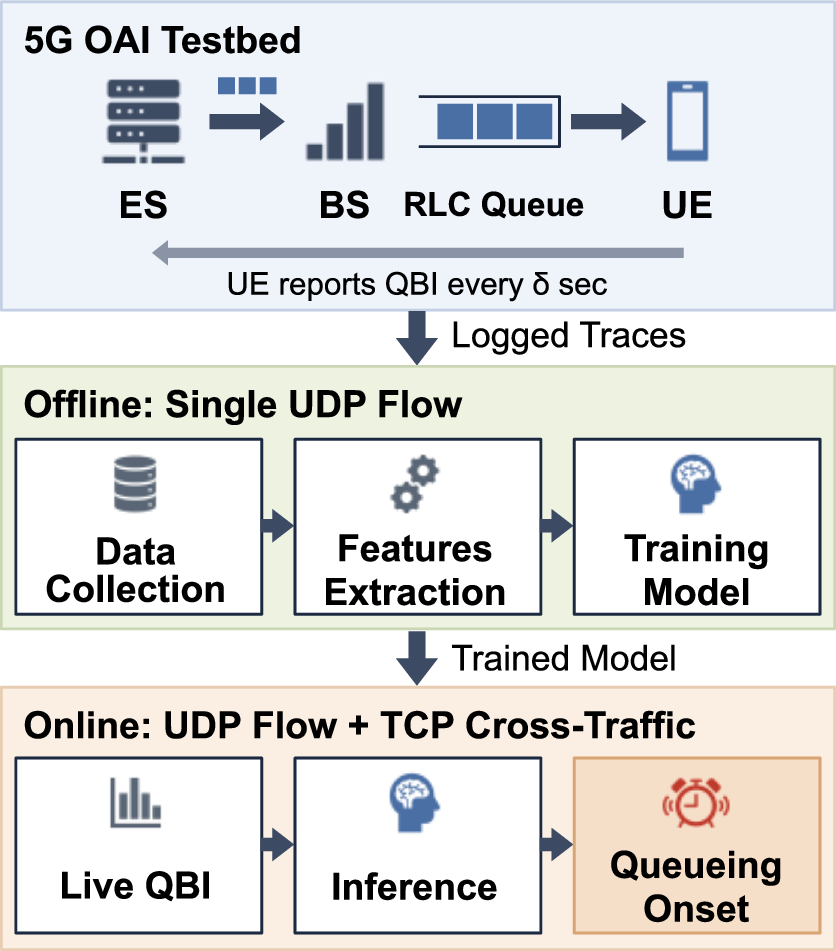}
    \caption{QueueTrack framework, including offline model training and online QBI-based inference.}
    \label{fig:pipeline}
    \vspace{-0.20cm}
\end{figure}

This section describes the input feature construction, classification models, and evaluation protocol used by QueueTrack to detect queue buildup episodes. 
Fig.~\ref{fig:pipeline} provides an overview of the QueueTrack framework, highlighting the offline training and online inference phases. 
\subsection{Feature Construction} \label{sec:feature_engineering}
Since queue buildup is a temporal phenomenon, QueueTrack does not rely on a single QBI sample. 
An isolated positive QBI value may arise from short-term scheduling variation, channel fluctuation, or measurement noise, whereas persistent positive QBI values over multiple reporting intervals indicate sustained queue buildup. 
Similarly, persistent negative QBI values indicate that the queue is being depleted. 

Let $\mathrm{QBI}_{i,j}$ denote the QBI value reported by UE $i$ during the $j$th feedback interval, as defined in Section~\ref{sec:metric}. 
We define $W$ as the number of consecutive reporting intervals, including the current interval, used for feature extraction. 
Using the current interval and the previous $W-1$ intervals, QueueTrack constructs an input feature vector. 

The feature vector combines, over the $W$ reporting intervals, (i) summary statistics of the recent QBI history and (ii) the ordered QBI sequence itself. 
The summary statistics include the mean QBI value $\mathrm{QBI}_{\mathrm{avg}}$, the fraction of positive QBI values $\mathrm{QBI}_{\mathrm{pos}}$, and the cumulative sum of negative QBI values $\mathrm{QBI}_{\mathrm{neg}}$. 
$\mathrm{QBI}_{\mathrm{avg}}$ captures the persistent trend of the queue to grow or drain over the observation period. 
$\mathrm{QBI}_{\mathrm{pos}}$ characterizes how often the estimated offered packet count $M_{i,j}$ exceeds the delivered packet count $N_{i,j}$ during the same observation window, thereby indicating queue buildup. 
$\mathrm{QBI}_{\mathrm{neg}}$ reflects how the queue is being depleted and makes it possible to distinguish a queue that is already draining from one that is starting to build up. 

Furthermore, QueueTrack includes the ordered QBI sequence $\{\mathrm{QBI}_{i,j}, \mathrm{QBI}_{i,j-1}, \ldots, \mathrm{QBI}_{i,j-W+1}\}$, which captures the recent temporal dynamics of the QBI trajectory. 
This sequence enables the classifiers to differentiate between persistent queue buildup, transient fluctuations, queue depletion, and recovery to baseline conditions. 
We also evaluated additional features, including the moving standard deviation, first-order differences, and exponentially weighted moving average of QBI. Since these features did not consistently improve performance, we use the feature set above to retain the key temporal structure while keeping the input representation compact. 
\subsection{Learning Models} \label{sec:learningmodels}
We evaluate QueueTrack using the following lightweight supervised classifiers implemented with the scikit-learn library~\cite{scikit-learn}: LR, SVM with a radial-basis-function (RBF) kernel, and MLP. 
These models are used to determine whether a fixed-size feature vector derived from recent QBI values is sufficient for classifying the queue state. 
LR serves as a linear baseline, whereas SVM and MLP are designed to capture nonlinear dependencies among the QBI summary statistics and ordered-sequence features. 
OWD is used solely to generate offline congestion labels and is excluded as an input feature during both training and inference. 
Unless otherwise stated, the models are trained on single-flow UDP traces and evaluated on held-out test sets that include both UDP traces and previously unseen mixed-flow traces. 
The model hyperparameters are tuned using the training split and then held fixed across all reported test scenarios. 
\HH{A fixed decision threshold of $0.5$ is applied to map the model outputs to binary decisions.} 
\subsection{Evaluation Protocol and Metrics} \label{sec:evalprotocol}
We report precision, recall, F1 score, and precision-recall (PR) curves because queue buildup detection is an imbalanced binary classification problem, where false alarms and missed detections have different operational consequences. 
Precision measures the fraction of predicted congestion alarms that are correct; low precision corresponds to an excess of unwarranted congestion responses. 
Recall captures the fraction of true congested intervals that are successfully identified; low recall indicates that the model fails to detect queue buildup episodes and may allow delay to grow unchecked. 
The F1 score, defined as the harmonic mean of precision and recall, is used as the primary scalar metric because it penalizes both false positives and undetected congestion events. 
PR curves complement the fixed-threshold F1 results by showing the trade-off between conservative high-confidence alarms and more aggressive detection settings. 
This is important because system designers may prefer different operating points depending on latency sensitivity and tolerance for false congestion signals. 
\section{Experimental Results} \label{sec:results}
This section evaluates whether QBI provides a practical short-history signal for detecting queue buildup at the network edge, rather than merely correlating with delay after congestion has already occurred. 
Since the RLC queue occupancy is typically unavailable to end hosts, we use congestion labels derived from OWD measurements as the ground truth. 
We conduct an in-depth evaluation using parameter sensitivity studies, out-of-regime generalization tests, precision-recall analysis, time-series inspection, and robustness evaluations under impaired UE feedback. 
The results show that QBI-derived features remain reliable across different feedback intervals $\delta$, window lengths $W$, previously unseen TCP Cubic and BBR cross-traffic settings, and moderate feedback corruption, supporting the viability of QBI as a practical queue-state signal for detecting the onset of queueing at the BS. 
\subsection{QBI Early-Warning Capability}
\begin{figure}[tbp]
    \centering
    \includegraphics[width=0.55\columnwidth]{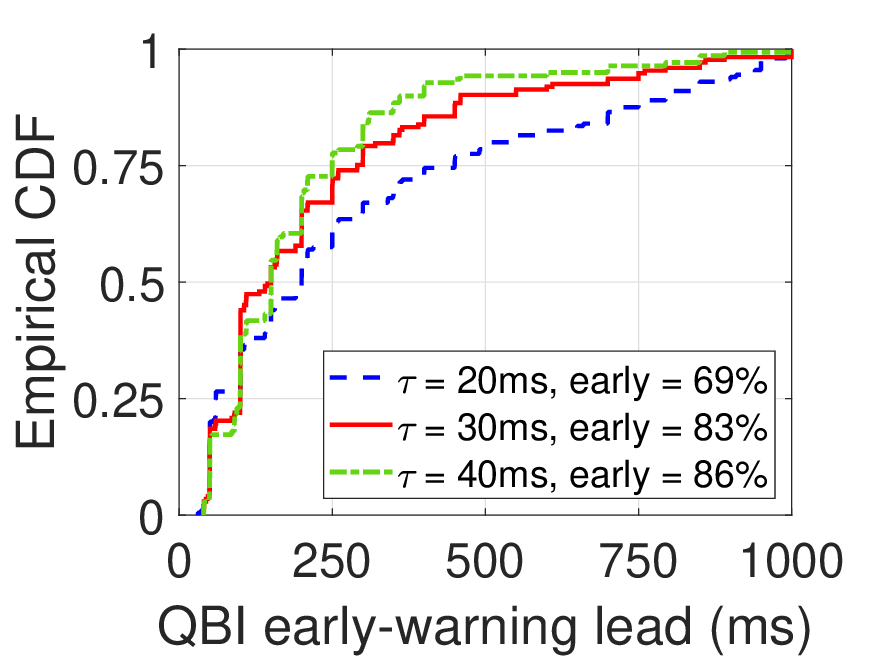}
    \caption{Empirical CDF of QBI early-warning lead times for different OWD thresholds. The percentages indicate the fraction of high-OWD episodes preceded by a persistent positive-QBI signal. Experimental results.}
    \label{fig:qbiearly}
    \vspace{-0.20cm}
\end{figure}
\begin{figure*}[tbp]
    \subfigure[F1 score vs. reporting interval $\delta$]{\includegraphics[width=0.53\columnwidth]{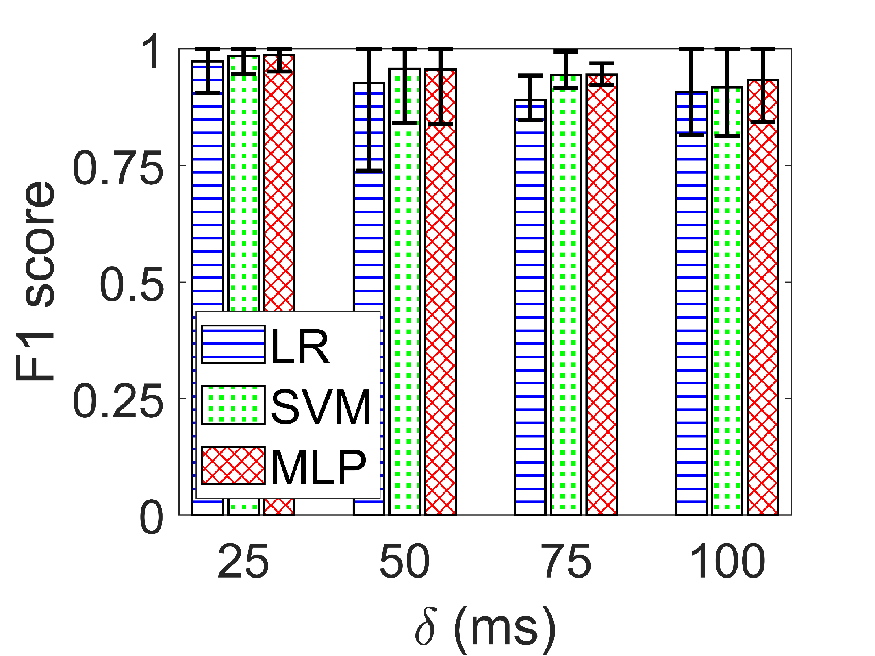}}
    \hspace{-20pt}
    \subfigure[F1 score vs. OWD threshold $\tau$]{
    \includegraphics[width=0.53\columnwidth]{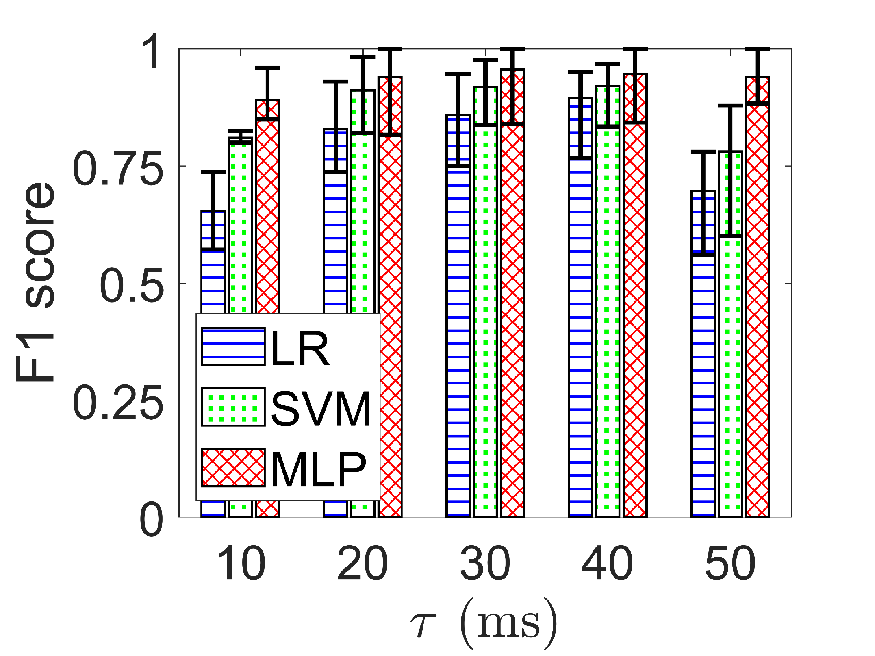}}
    \hspace{-20pt}
    \subfigure[F1 score vs. window size $W$]{
    \includegraphics[width=0.53\columnwidth]{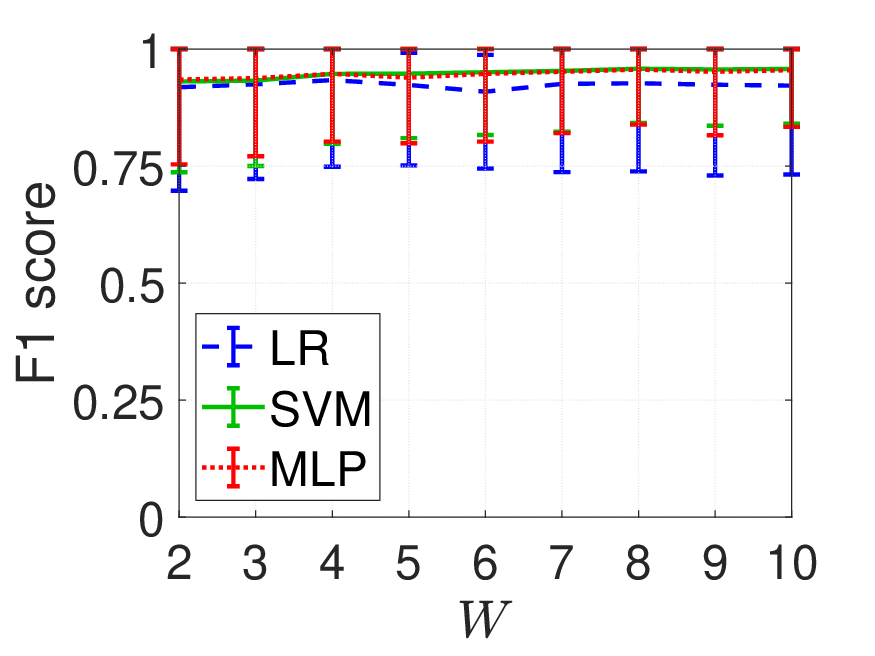}}
    \hspace{-19pt}
    \subfigure[Feature importance heatmap]{
    \includegraphics[width=0.53\columnwidth]{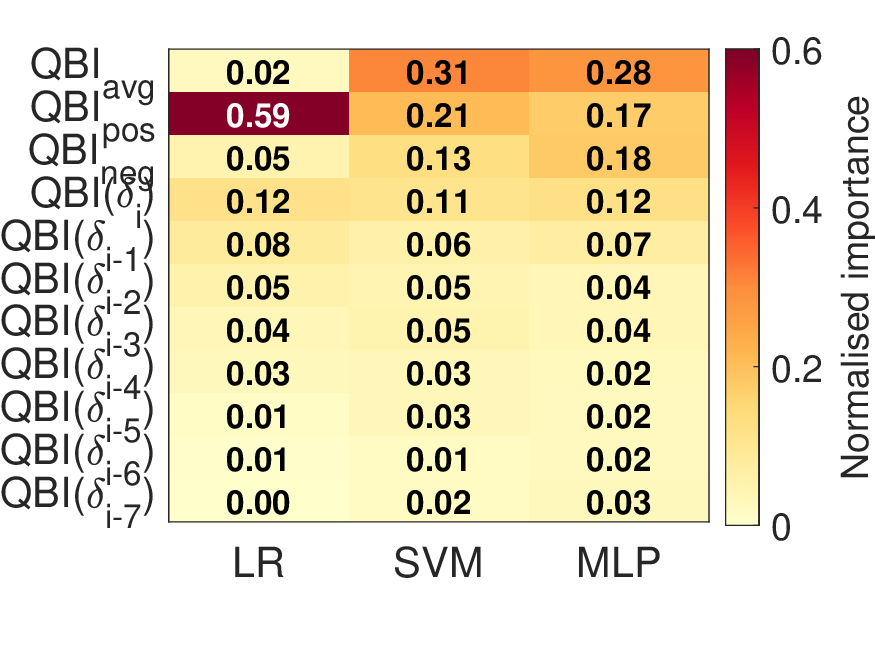}}
    \hspace{-20pt}
    \caption{Single-flow UDP sensitivity and feature analysis, illustrating the impact of (a) the reporting interval $\delta$, (b) the OWD threshold $\tau$, and (c) the history window size $W$ on detection performance, and (d) the importance of QBI-derived input features.}
    \vspace{-0.20cm}
    \label{fig:single_UDP}
\end{figure*}
\begin{figure}[tbp]
    \centering
    \includegraphics[width=0.55\columnwidth]{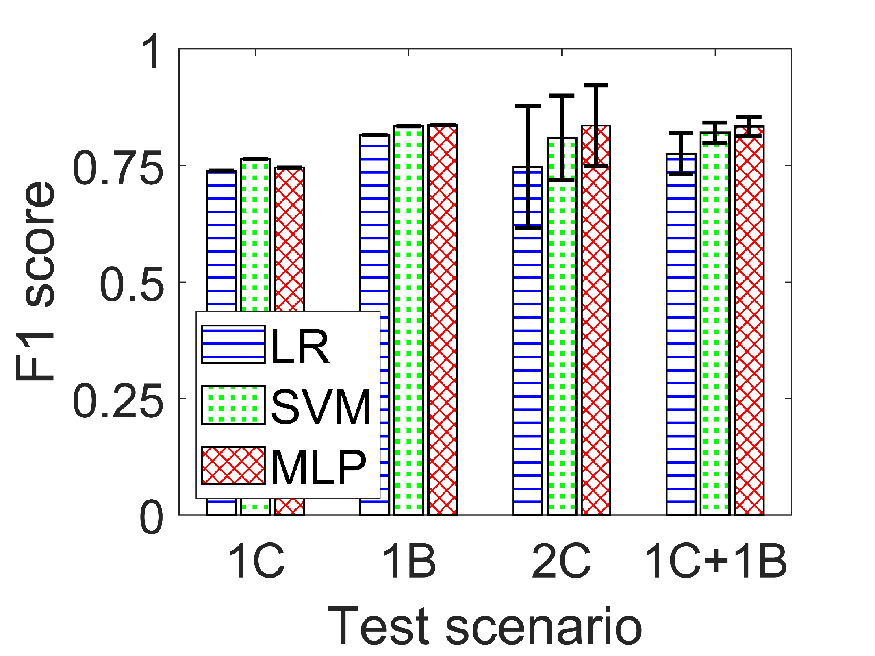}
    \caption{F1 scores under unseen mixed cross-traffic scenarios.}
    \label{fig:mixedflows_f1}
    \vspace{-0.20cm}
\end{figure}

\HH{The behavior shown in Fig.~\ref{fig:qbi} indicates that QBI can reveal queue-state changes before the corresponding changes are reflected in OWD. 
To assess the early-warning capability of QBI, we define a persistent high-OWD episode as a sequence of at least $K$ consecutive feedback reports with $\text{OWD} > \tau$, and a persistent positive-QBI period as a sequence of at least $K$ consecutive feedback reports with $\text{QBI} > 0$. 
We set $K=2$ empirically to filter out isolated fluctuations while preserving short-term queue changes. 
For each episode, we determine the most recent persistent positive-QBI period that appears before OWD exceeds this threshold. 
We then define the lead time as the time interval between the start of this positive-QBI period and the subsequent crossing of the OWD threshold.} 

Fig.~\ref{fig:qbiearly} presents the empirical CDF of the resulting lead times. 
We adopt $\tau = 30$ms as the default threshold: $\tau = 20$ms tends to label even small delay increases as congestion, while $\tau = 40$ms detects only more severe events and therefore yields fewer positive episodes. 
The sensitivity of the classifiers to the choice of $\tau$ is evaluated separately using the F1 scores in Section~\ref{sec:singleflow}. 
With $\tau = 30$ms, $173$ out of $209$ OWD episodes ($83$\%) were preceded by a persistent positive-QBI signal. 
Across these $173$ episodes, the median lead time was $149.8$ms (interquartile range of $99.8$-$290.6$ms), while the mean lead time was $225.7$ms, with a $95$\% confidence interval of $193.7$-$258.9$ms. 
For the remaining episodes, a persistent positive-QBI signal either occurred only after the OWD threshold crossing or was not observed. 
\subsection{QBI-Based Detection in a Single-Flow UDP Scenario} \label{sec:singleflow}
We evaluate detection performance in a single-flow UDP setting to isolate the impact of key design parameters. 
Specifically, we examine the UE reporting interval $\delta$, the history window length $W$, and the OWD congestion threshold $\tau$. 
The results are summarized in Fig.~\ref{fig:single_UDP}. 
Unless varied on the x-axis, we use the default values of $\delta = 50$ms, $W = 8$, and $\tau = 30$ms. 

Fig.~\ref{fig:single_UDP} (a) presents the F1 score as the reporting interval $\delta$ is varied. 
The models maintain high F1 scores over the tested range, indicating that QueueTrack does not depend on any specific feedback interval. 
Each bar’s height represents the aggregate F1 score across the four single-flow rate settings in $\{10, 20, 30, 40\}$Mbps, while the vertical ranges indicate the minimum and maximum F1 values observed at these individual rates. 
These ranges therefore capture variation across load conditions rather than statistical uncertainty. 
Their limited spread further indicates that performance is stable across the tested single-flow rates. 
Since performance remains strong over the considered range, we adopt $\delta = 50$ms for subsequent experiments as a representative operating point. 
This choice reflects a practical trade-off: a smaller $\delta$ increases the frequency of UE reports and thus uplink reporting overhead, whereas a larger $\delta$ reduces overhead at the cost of coarser temporal resolution. 

Fig.~\ref{fig:single_UDP} (b) analyzes how sensitive QueueTrack is to the choice of OWD threshold $\tau$, where samples are labeled as congested if $\text{OWD} > \tau$. 
The threshold $\tau$ directly determines which samples are considered positive and how frequently positive samples occur. 
Smaller values of $\tau$ mark even modest delay increases as congestion, which diminishes the separation between classes. 
Larger values restrict the positive class to pronounced delay spikes, thereby increasing class imbalance. 
Over the range of thresholds considered, the models consistently achieve high F1 scores, with the best results occurring at intermediate thresholds. 
Since performance is close to its maximum for $\tau$ around $20$-$30$ms, we select $\tau = 30$ms in the subsequent experiments. 
Relative to $\tau=20$ms, this choice yields a slightly stricter notion of congestion by excluding samples with only slight delay increases, while still avoiding the scarcity of positive samples seen at higher thresholds. 
In practical deployments, $\tau$ should be viewed as a tunable, system- or service-specific parameter that dictates when elevated OWD is classified as congestion. 

Fig.~\ref{fig:single_UDP} (c) shows the effect of the history window length $W$. 
Detection accuracy improves once $W$ exceeds the shortest histories, indicating that the congestion is more reliably inferred from recent temporal evolution than from near-instantaneous measurements alone. 
The gains then diminish as $W$ increases further, with performance saturating at modest window lengths. 
This suggests that the most informative part of the QBI signal lies in the immediate past, while extending the temporal memory offers only marginal additional gains. 

Finally, Fig.~\ref{fig:single_UDP} (d) shows the normalized importance of the QBI-based input features. 
The heatmap indicates that the models place the greatest emphasis on compact summary statistics and recent QBI history, rather than relying primarily on more distant QBI values. 
\subsection{Generalization to Unseen Mixed-Flow Scenarios}
We investigate whether QBI-based detection can generalize beyond the single-flow UDP training setup. 
To this end, we analyze the behavior of QueueTrack under varying cross-traffic conditions. 
The models are trained exclusively on single-flow UDP traces and then evaluated on scenarios with previously unseen mixed cross-traffic, as follows: one TCP Cubic flow (1C), one TCP BBR flow (1B), two TCP Cubic flows (2C), and mixed TCP Cubic and BBR flows (1C+1B)\HH{, see Table~\ref{tab:multiue}.} 
These scenarios match the multi-UE setup described in Section~\ref{sec:measurement}. 

Fig.~\ref{fig:mixedflows_f1} reports the F1 scores for these settings, showing that the models maintain strong detection performance across all previously unseen scenarios. 
LR achieves F1 scores of about $0.74$–$0.82$, while the nonlinear SVM and MLP classifiers improve performance to roughly $0.76$–$0.84$. 
The best results are obtained under TCP BBR cross-traffic (1B), where SVM and MLP both achieve F1 scores of $0.83$. 
The scenarios with only TCP Cubic cross-traffic are more challenging: the 1C case yields the lowest scores for all classifiers, and the 2C setting exhibits larger variability across loads, indicating that TCP Cubic induces queue dynamics that are less clearly distinguishable in the QBI feature space. 
Nevertheless, both SVM and MLP retain F1 scores above $0.80$ in 2C and 1C+1B, demonstrating that QBI-based features capture a congestion-state signal that generalizes well to previously unseen cross-traffic conditions, rather than simply overfitting to the single-flow UDP training distribution.
\begin{figure}[tbp]
    \centering
    \includegraphics[width=0.75\columnwidth]{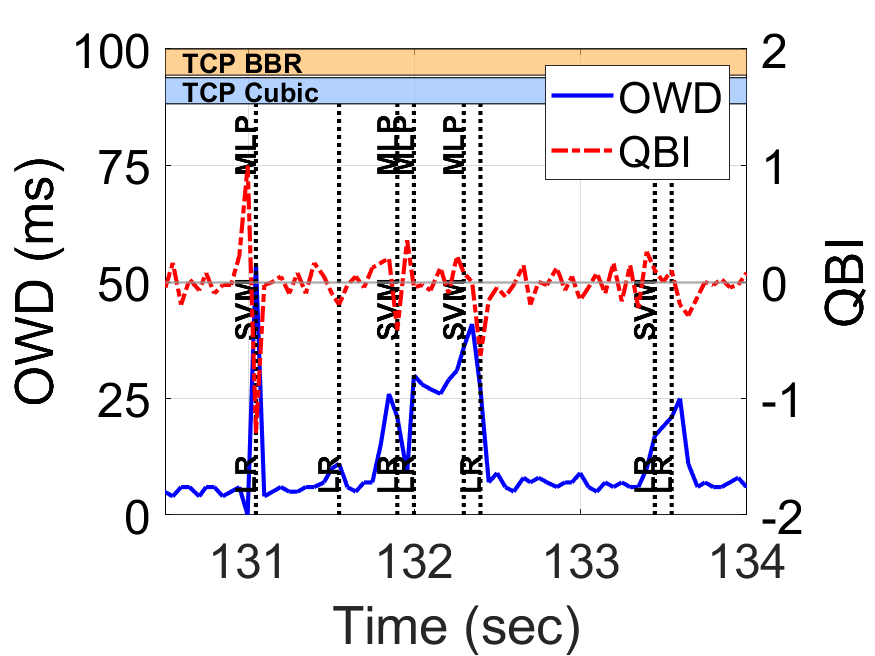}
    \caption{Representative time-series comparison of model decisions.}
    \label{fig:mixedflows_trace}
    \vspace{-0.20cm}
\end{figure}

Fig.~\ref{fig:mixedflows_trace} illustrates a time-series performance comparison of QueueTrack with LR, SVM, and MLP models in detecting the onset of queueing for a representative 5G testbed trace. 
In this experiment, the UDP flow runs at $10$Mbps, while two concurrent TCP flows, i.e., one TCP Cubic and one TCP BBR, serve as cross-traffic (1C+1B). 
LR identifies the congestion episodes, but also produces false positives at $t=131.5$sec and $t=133.5$sec. 
The nonlinear SVM and MLP models track the onset of OWD increases more tightly, indicating that the relationship between QBI dynamics and the queueing state is not strictly linear. 
Within this short interval, the MLP achieves slightly higher accuracy. For example, at $t=132$sec, the MLP correctly identifies the onset of queueing, whereas the SVM misses it. 
Nonetheless, all three models exhibit a false positive around $t=133.5$sec. 
For the $90$sec interval in which all three flows are active, the F1 scores of LR, SVM, and MLP are $0.71$, $0.77$, and $0.78$, respectively. 
\subsection{Precision-Recall Analysis}
\begin{figure*}[tbp]
    \subfigure[1 TCP Cubic flow (1C)]{   
    \includegraphics[width=0.53\columnwidth]{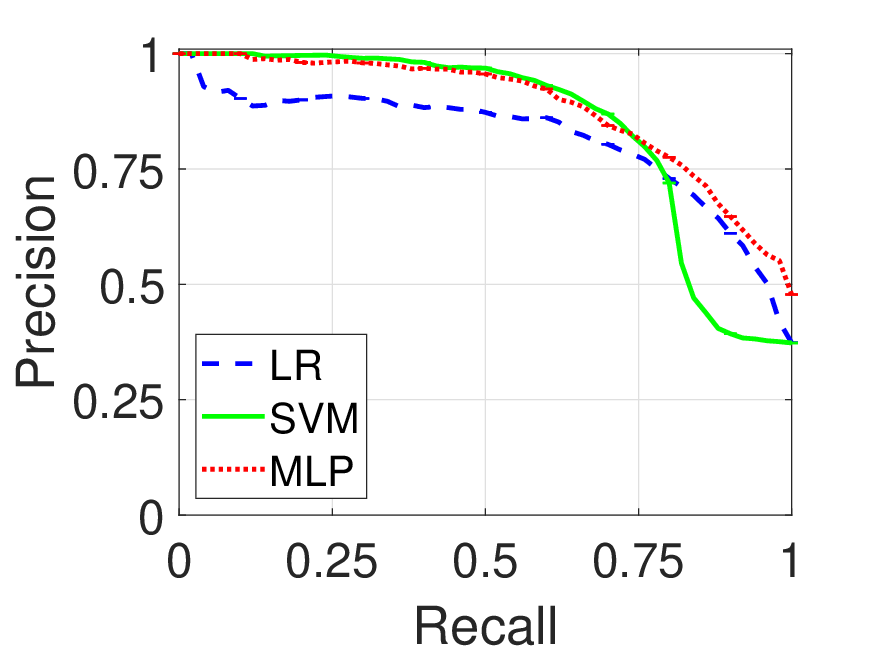}} 
    \hspace{-20pt}
    \subfigure[1 TCP BBR flow (1B)]{
    \includegraphics[width=0.53\columnwidth]{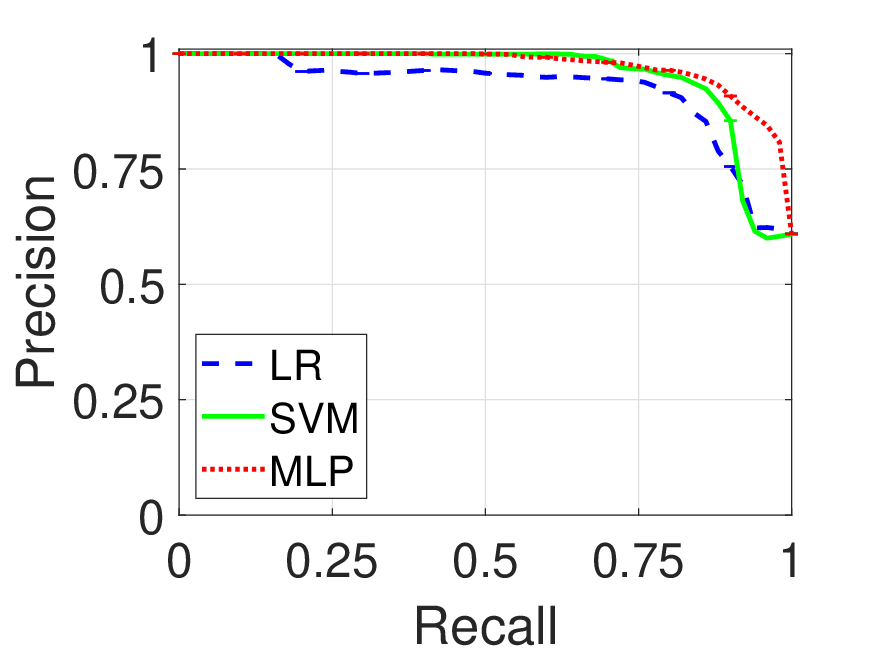}}
    \hspace{-20pt}
    \subfigure[2 TCP Cubic flows (2C)]{
    \includegraphics[width=0.53\columnwidth]{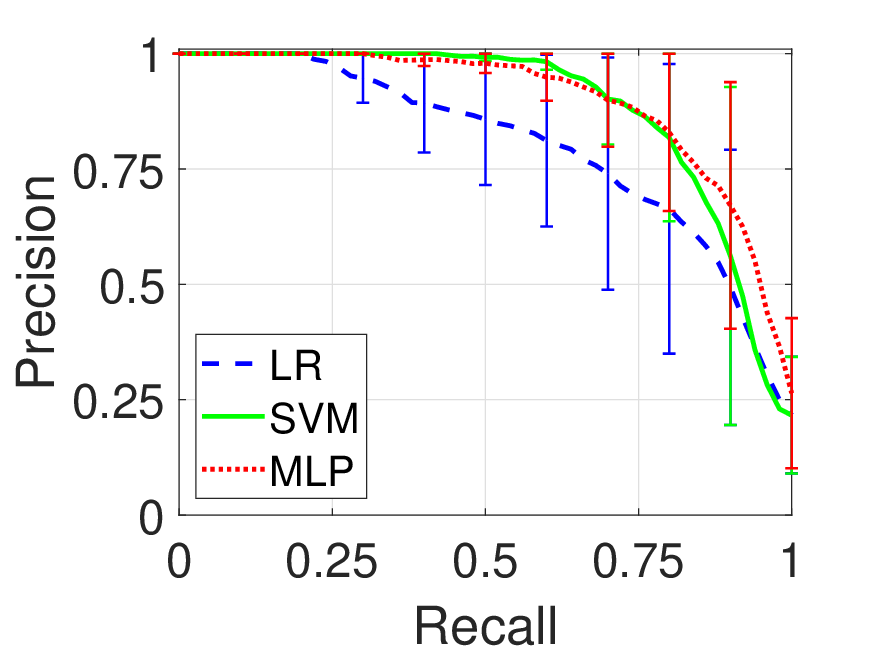}}
    \hspace{-20pt}
    \subfigure[TCP Cubic and BBR flows (1C+1B)]{
    \includegraphics[width=0.53\columnwidth]{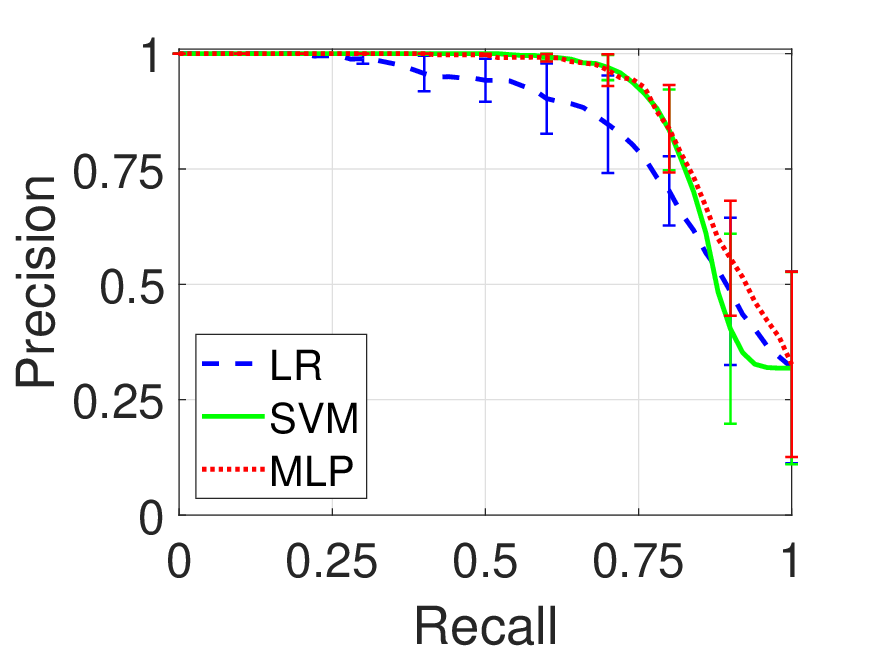}}
    \hspace{-20pt}
    \caption{Precision-recall performance under different cross-traffic scenarios for LR, SVM, and MLP classifiers.}
    \label{fig:mixed_PR}
    \vspace{-0.20cm}
\end{figure*}

The PR curves in Fig.~\ref{fig:mixed_PR} provide a more detailed view of generalization than the aggregate F1 scores alone. 
Across unseen mixed-flow scenarios, the QBI-based detectors maintain precision and recall, avoiding the shift toward low precision and high recall that would suggest overfitting to the single-flow UDP training configuration. 
Moreover, SVM and MLP preserve a broad operating range where precision remains high as recall increases, indicating that QBI captures congestion-state structure that remains stable under varying cross-traffic conditions. 
This behavior is most pronounced for TCP BBR (1B), see Fig.~\ref{fig:mixed_PR} (b), where the nonlinear detectors maintain near-perfect precision over much of the recall range. 
This suggests that QBI provides a strong representation of the queue dynamics induced by TCP BBR. 

The scenarios with TCP Cubic cross-traffic are more challenging, see, e.g., Fig.~\ref{fig:mixed_PR} (a). 
In these cases, the curves bend downward earlier, particularly at high recall, indicating that the detection of more congestion episodes comes at the cost of a higher false positive rate. 
Notably, in the 2C (Fig.~\ref{fig:mixed_PR} (c)) and 1C+1B (Fig.~\ref{fig:mixed_PR} (d)) scenarios, the curves degrade smoothly rather than abruptly, showing that the detector remains effective as the number of flows increases and the cross-traffic becomes more heterogeneous. 
From a deployment perspective, QBI does not impose a single fixed operating point: system designers can select a conservative threshold for high-confidence congestion alerts or relax the threshold when broader congestion coverage is preferred. 
The stable precision–recall tradeoff observed under previously unseen cross-traffic conditions shows that QBI offers a reliable signal for the onset of queueing at the BS when the cross-traffic is not known in advance. 
\subsection{Robustness Under Feedback Impairments}
\begin{figure}[t]
    \centering
    \begin{minipage}[t]{0.50\columnwidth}
        \centering
        \subfigure[F1]{%
            \includegraphics[width=\linewidth]{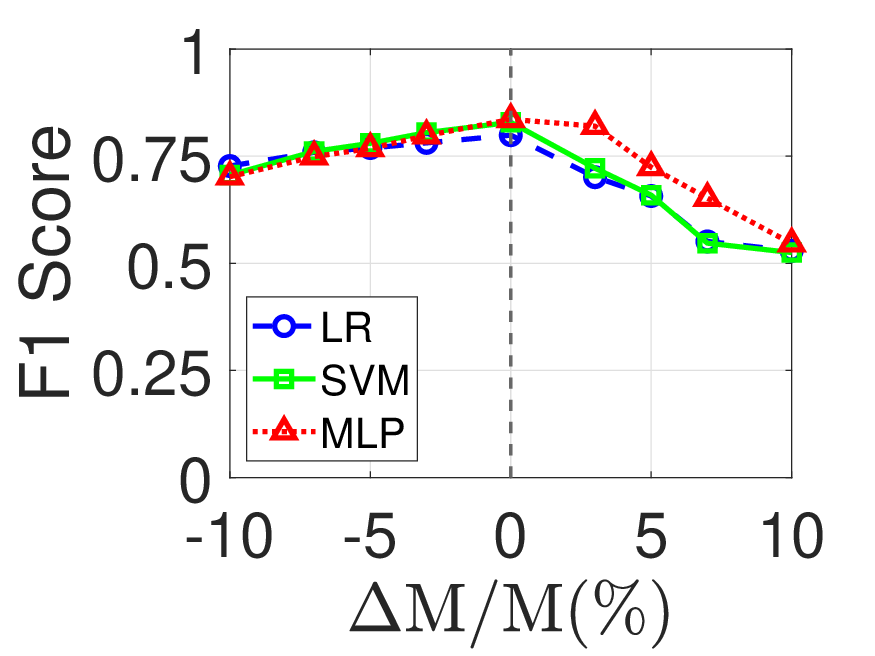}}
    \end{minipage}\hfill
    \begin{minipage}[t]{0.50\columnwidth}
        \centering
        \subfigure[Precision]{%
            \includegraphics[width=\linewidth]{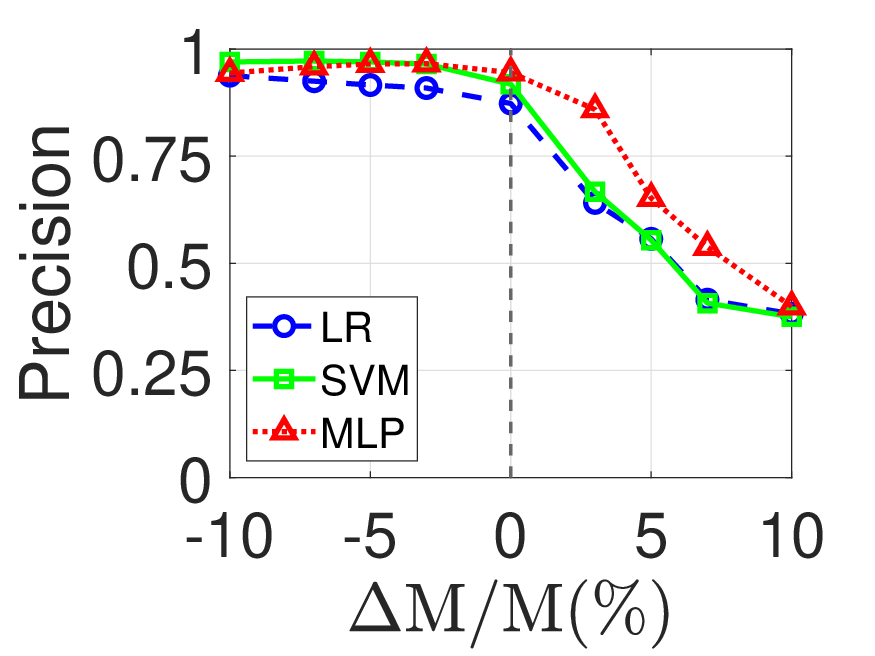}}
    \end{minipage}
    \vspace{1mm}
    \begin{minipage}[t]{0.50\columnwidth}
        \centering
        \subfigure[Recall]{%
            \includegraphics[width=\linewidth]{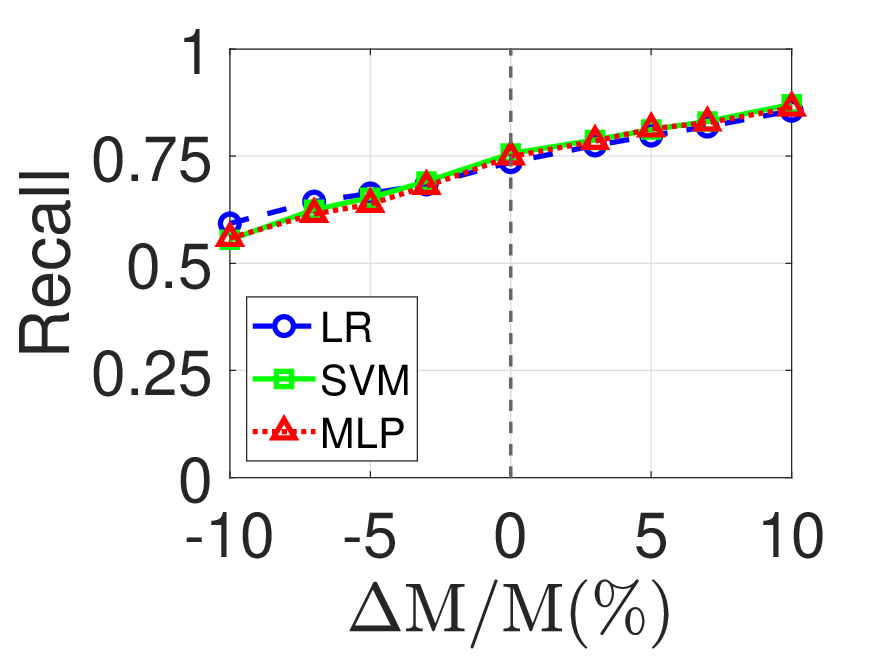}}
    \end{minipage}\hfill
    \begin{minipage}[t]{0.50\columnwidth}
        \centering
        \subfigure[F1]{%
            \includegraphics[width=\linewidth]{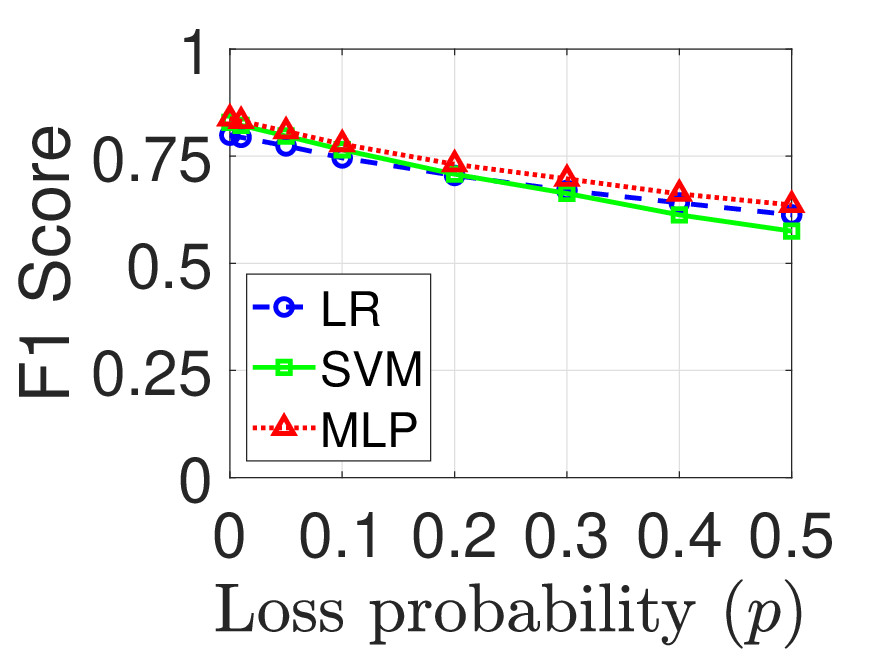}}
    \end{minipage}
    \vspace{1mm}
    \begin{minipage}[t]{0.50\columnwidth}
        \centering
        \subfigure[Precision]{%
            \includegraphics[width=\linewidth]{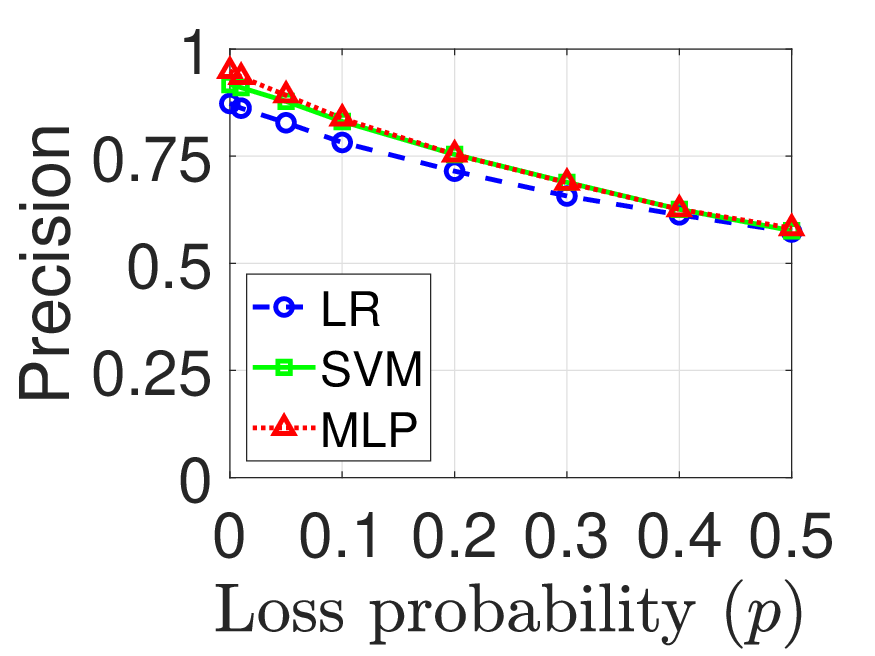}}
    \end{minipage}\hfill
    \begin{minipage}[t]{0.50\columnwidth}
        \centering
        \subfigure[Recall]{%
            \includegraphics[width=\linewidth]{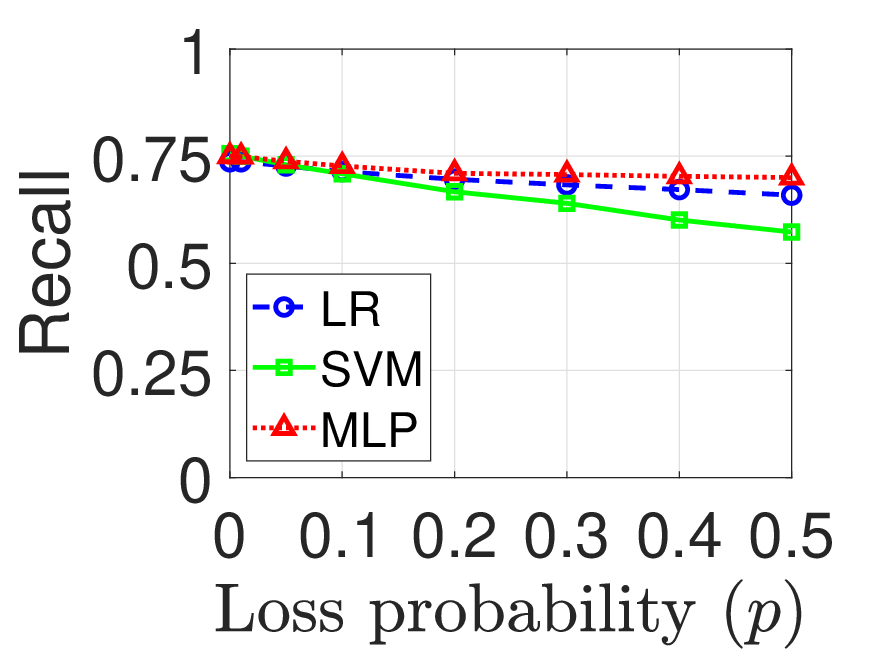}}
    \end{minipage}
    \caption{Robustness under impaired UE feedback. Panels (a)--(c) inject multiplicative error into the offered packet count $M$, while panels (d)--(f) independently drop feedback reports with Bernoulli probability $p$.}
    \vspace{-0.20cm}
    \label{fig:robustness_combined}
\end{figure}

QueueTrack relies on periodic UE feedback to reconstruct QBI-based features. 
However, in practice, these feedback reports may be imperfect due to uplink packet loss or measurement noise. 
For instance, the QBI computation assumes that the send rate remains constant within each reporting interval $\delta$. 
This assumption can be violated by rate adaptation at the transport layer, operating-system scheduling delays, or other sources of timing noise. 
We evaluate the robustness of QueueTrack under two feedback-corruption scenarios. 
First, we introduce a relative error into $M$ at each feedback interval, with $\Delta M/M \in \{-10,-7,-5,-3,0,3,5,7,10\}\%$. 
For each corrupted value of $M$, we recompute QBI using Eq.~\eqref{eq:qbi_definition} and rebuild the input features based on the resulting corrupted QBI sequence. 
Second, we model independent feedback loss by dropping each feedback report with Bernoulli probability $p \in \{0,0.01,0.05,0.10,0.20,0.30,0.40,0.50\}$. 
Dropped reports are imputed using the most recent UE feedback sample. Fig.~\ref{fig:robustness_combined} reports the resulting F1 score, precision, and recall.

In the relative-error analysis of $M$, QueueTrack is more sensitive to overestimation than to underestimation. 
As illustrated in Fig.~\ref{fig:robustness_combined} (b), precision drops for positive $\Delta M / M$, indicating that QueueTrack predicts congestion more aggressively and generates additional false positives. 
Negative perturbations have the opposite effect: QueueTrack behaves more conservatively and predicts congestion less often, increasing the number of false negatives and reducing recall, as shown in Fig.~\ref{fig:robustness_combined} (c). 
The MLP is more tolerant to moderate positive bias than LR and SVM.  
As depicted in Fig. \ref{fig:robustness_combined} (a), at $\Delta M/M = 3\%$, the MLP F1 score remains close to its unperturbed value ($\approx 0.82$ versus $\approx 0.84$), while LR and SVM drop to approximately $0.70$ and $0.72$, respectively. 
At $\Delta M/M = 5\%$, all models degrade, with F1 scores of roughly $0.66$, $0.66$, and $0.72$ for LR, SVM, and MLP, respectively. 
Overall, positive error in $M$ is more harmful than comparable negative error because it biases the reconstructed QBI sequence toward apparent backlog growth, increasing false positives. 

In contrast, feedback loss leads to a gradual decline in F1 scores with substantial degradation emerging only when $p > 0.20$, see Fig.~\ref{fig:robustness_combined} (d). 
As $p$ increases, precision drops more sharply than recall, see Fig.~\ref{fig:robustness_combined} (e) and (f), because the observation window contains fewer up-to-date feedback measurements, causing outdated values to be propagated over longer intervals. 
This temporal mismatch can make the inferred queue dynamics appear more congested than they actually are, leading to more false positives and consequently reducing precision. 
\section{Conclusion} \label{sec:conclusion}
This paper presented QueueTrack, a practical end-host framework that infers bottleneck queue dynamics using endpoint-observable measurements to detect the onset of queueing in mobile networks. 
QueueTrack computes a queue buildup indicator, denoted QBI, from packet-level timing statistics and uses finite-history QBI features with low-complexity learning models to identify queue-state transitions at the RLC layer of the BS. 
The framework operates without requiring tight synchronization between end hosts or modifications to existing network infrastructure or 3GPP standards. 
We implemented and evaluated the complete QueueTrack ML pipeline on an OAI 5G testbed. 
The evaluation covered single-flow UDP settings, multi-UE scenarios with previously unseen TCP Cubic and BBR cross-traffic, and impaired feedback conditions. 
The results show that QBI provides an effective signal for detecting the onset of queueing. 
Under challenging unseen traffic scenarios, QueueTrack maintains reliable detection performance, with nonlinear models such as SVM and MLP achieving F1 scores around $80\%$. 
The robustness evaluation further shows that QueueTrack is resilient to moderate feedback loss, while positive bias in the estimated offered packet count can increase false positives. 
The experimental results demonstrate that QueueTrack can support low-latency CCAs, as well as network telemetry, runtime diagnostics, and network monitoring in mobile networks. 
We extend QueueTrack from queueing-onset detection to near-future queueing-onset prediction using an echo state network~\cite{Karnam26}. 
As future work, we intend to evaluate the robustness of QBI and QueueTrack under different MAC scheduler configurations, scheduler hyperparameter settings, and mobility conditions. 
\section*{Acknowledgment}
This work was supported by the DAIS project under Grant Agreement No. 101007273, the HICONNECTS project under Grant Agreement No. 101097296, and the Dutch RVO under TSH project No. 21007. 
\bibliographystyle{IEEEtran}
\bibliography{references}

@article{Saad20,
    author={Saad, Walid and Bennis, Mehdi and Chen, Mingzhe},
    journal={IEEE Network}, 
    title={{A Vision of 6G Wireless Systems: Applications, Trends, Technologies, and Open Research Problems}}, 
    year={2020},
    volume={34},
    number={3},
    pages={134-142}}

@article{Hassani21,
    author={Hassani, Hamid and Gringoli, Francesco and Leith, Douglas J.},
    journal={Computer Networks},
    pages={107820},
    title={{Quick \& Plenty: Achieving Low Delay \& High Rate in 802.11ac Edge Networks}},
    volume={187},
    year={2021}}

@article{Ha08,
    author={Ha, Sangtae and Rhee, Injong and Xu, Lisong},
    title={{CUBIC: A New TCP-Friendly High-Speed TCP Variant}},
    year={2008},
    volume={42},
    number={5},
    journal={ACM SIGOPS Operating Systems Review},
    pages={64-74},
    month={July}}

@article{Xie22,
    author={Xie, Yaxiong and Jamieson, Kyle},
    journal={Proceedings of the ACM on Measurement and Analysis of Computing Systems},
    title={{NG-Scope: Fine-Grained Telemetry for NextG Cellular Networks}},
    year={2022},
    volume={6},
    number={1},
    pages={1-26},
    month={Feb}}

@article{She23,
    author={She, Changyang and Duong, Trung Q. and Quek, Tony Q. S. and Viswanathan, Harish and Lopez-Perez, David},
    journal={IEEE Wireless Communications}, 
    title={{Guest Editorial: Intelligent Ultra-Reliable and Low-Latency Communications in 6G}}, 
    year={2023},
    volume={30},
    number={2},
    pages={12-13}}

@article{Akyildiz22,
    title={{Wireless Communication Research Challenges for Extended Reality (XR)}},
    author={Akyildiz, Ian F. and Guo, Hongzhi},
    journal={ITU Journal on Future and Evolving Technologies},
    year={2022},
    volume={3}, 
    number={2},
    pages={273-287}}

@article{BBR,
    author={Cardwell, Neal and Cheng, Yuchung and Gunn, C. Stephen and Hassas Yeganeh, Soheil and Jacobson, Van},
    title={{BBR: Congestion-Based Congestion Control}},
    journal={Communications of the ACM},
    year={2017},
    month={Jan},
    volume={60},
    number={2},
    pages={58-66}}

@book{Tanenbaum21,
    title={{Computer Networks}},
    author={Tanenbaum, Andrew S. and Feamster, Nick and Wetherall, David J.},
    year={2021},
    edition={6},
    publisher={Pearson}}

@misc{RFC9330,
    author={Briscoe, Bob and De Schepper, Koen and Bagnulo, Marcelo and White, Greg},
    title={{Low Latency, Low Loss, and Scalable Throughput (L4S) Internet Service: Architecture}},
    series={Request for Comments},
    number={9330},
    howpublished={RFC 9330},
    publisher={RFC Editor},
    year={2023},
    month={Jan}}

@inproceedings{ExLL18,
    author={Park, Shinik and Lee, Jinsung and Kim, Junseon and Lee, Jihoon and Ha, Sangtae and Lee, Kyunghan},
    title={{ExLL: An Extremely Low-Latency Congestion Control for Mobile Cellular Networks}},
    booktitle={Proceedings of the 14th International Conference on Emerging Networking EXperiments and Technologies (CoNEXT)},
    year={2018},
    pages={307-319},
    location={Heraklion, Greece}}

@inproceedings{PBE20,
    title={{PBE-CC: Congestion Control via Endpoint-Centric, Physical-Layer Bandwidth Measurements}},
    author={Xie, Yaxiong and Yi, Fan and Jamieson, Kyle},
    booktitle={Proceedings of the ACM SIGCOMM Conference},
    pages={451-464},
    month={Aug},
    year={2020}}

@techreport{Adeia23,
    type={{White Paper}},
    author={{Adeia}},
    title={{Immersive Media: Bringing Science Fiction to Commercial Reality}},
    year={2023},
    month={May},
    institution={Adeia}}

@techreport{SK23,
    type={{White Paper}},
    author={{SK Telecom}},
    title={{5G Lessons Learned, 6G Key Requirements, 6G Network Evolution, and 6G Spectrum}},
    year={2023},
    month={Aug},
    institution={SK Telecom}}

@phdthesis{Flavien22,
    author={Ronteix--Jacquet, Flavien},
    title={{Reducing Latency and Jitter in 5G Radio Access Networks}},
    school={{Ecole nationale sup{\'e}rieure Mines-T{\'e}l{\'e}com Atlantique}},
    year={2022}}

@inproceedings{piStream15,
    author={Xie, Xiufeng and Zhang, Xinyu and Kumar, Swarun and Li, Li Erran},
    title={{piStream: Physical Layer Informed Adaptive Video Streaming over LTE}},
    booktitle={Proceedings of the 21st Annual International Conference on Mobile Computing and Networking (MobiCom)},
    pages={413-425},
    year={2015},
    location={Paris, France}}

@inproceedings{claw17,
    author={Xie, Xiufeng and Zhang, Xinyu and Zhu, Shilin},
    title={{Accelerating Mobile Web Loading Using Cellular Link Information}},
    booktitle={Proceedings of the 15th Annual International Conference on Mobile Systems, Applications, and Services (MobiSys)},
    pages={427-439},
    year={2017},
    location={Niagara Falls, New York, USA}}

@inproceedings{Pathak08,
    author={Pathak, Abhinav and Pucha, Himabindu and Zhang, Ying and Hu, Y. Charlie and Mao, Z. Morley},
    title={{A Measurement Study of Internet Delay Asymmetry}},
    booktitle={Proceedings of the 9th International Conference on Passive and Active Network Measurement (PAM)},
    pages={182-191},
    year={2008},
    location={Berlin, Heidelberg}}

@inproceedings{Xu04,
    author={Xu, Lisong and Harfoush, K. and Rhee, Injong},
    title={{Binary Increase Congestion Control (BIC) for Fast Long-Distance Networks}}, 
    booktitle={Proceedings of the 23rd Annual Joint Conference of the IEEE Computer and Communications Societies (INFOCOM)}, 
    volume={4},
    pages={2514-2524},
    year={2004},
    location={Hong Kong, China}}

@article{OAI,
    author={Nikaein, Navid and Marina, Mahesh K. and Manickam, Saravana and Dawson, Alex and Knopp, Raymond and Bonnet, Christian},
    title={{OpenAirInterface: A Flexible Platform for 5G Research}},
    journal={ACM SIGCOMM Computer Communication Review},
    pages={33-38},
    volume={44},
    number={5},
    month={Oct},
    year={2014}}

@article{Chefrour21,
    author={Chefrour, Djalel},
    journal={ACM Computing Surveys},
    title={{One-Way Delay Measurement From Traditional Networks to SDN: A Survey}},
    year={2021},
    volume={54},
    number={7},
    pages={1-35},
    month={July}}

@inproceedings{Wan24a,
    author={Wan, Haoran and Cao, Xuyang and Marder, Alexander and Jamieson, Kyle},
    title={{NR-Scope: A Practical 5G Standalone Telemetry Tool}},
    booktitle={Proceedings of the 20th International Conference on Emerging Networking EXperiments and Technologies (CoNEXT)},
    pages={73-80},
    location={Los Angeles, CA, USA},
    year={2024}}

@misc{Wan24b,
    title={{Evolving Mobile Cloud Gaming with 5G Standalone Network Telemetry}}, 
    author={Wan, Haoran and Jamieson, Kyle},
    year={2024},
    eprint={2402.04454},
    archivePrefix={arXiv},
    primaryClass={cs.NI},
    url={https://arxiv.org/abs/2402.04454}}

@article{Irazabal24,
    author={Irazabal, Mikel and Nikaein, Navid},
    journal={IEEE Journal on Selected Areas in Communications}, 
    title={{TC-RAN: A Programmable Traffic Control Service Model for 5G/6G SD-RAN}}, 
    year={2024},
    volume={42},
    number={2},
    pages={406-419}}

@inproceedings{Vleeschauwer24,
    author={De Vleeschauwer, Danny and Wassing, Tom J. and De Schepper, Koen and Chang, Chia-Yu and Papagianni, Chrysa},
    title={{Application-Driven Traffic Differentiation}}, 
    booktitle={Proceedings of the Joint European Conference on Networks and Communications \& 6G Summit (EuCNC/6G Summit)}, 
    year={2024},
    pages={1049-1054}}

@online{ptp4l,
    title={{ptp4l(8) - Linux Man Page}},
    url={https://linux.die.net/man/8/ptp4l},
    year={2025},
    note={Accessed: May 10, 2026}}

@online{phc2sys,
    title={{phc2sys(8) - Linux Man Page}},
    url={https://linux.die.net/man/8/phc2sys},
    year={2025},
    note={Accessed: May 10, 2026}}

@online{tcp,
    title={{tcp(7) — Linux Manual Page}},
    url={https://man7.org/linux/man-pages/man7/tcp.7.html},
    year={2025},
    note={Accessed: May 10, 2026}}

@inproceedings{Shin25,
    author={Shin, Juhun and Lee, Goodsol and Paek, Jeongyeup and Bahk, Saewoong},
    booktitle={Proceedings of the IEEE Conference on Computer Communications (INFOCOM)}, 
    title={{C{\'e}sar: Cellular Resource Scheduling-Aware Congestion Control}}, 
    year={2025},
    month={May},
    pages={1-10},
    location={London, United Kingdom}}

@misc{RFC3168,
    author={Ramakrishnan, K. K. and Floyd, Sally and Black, David L.},
    title={{The Addition of Explicit Congestion Notification (ECN) to IP}},
    series={Request for Comments},
    number={3168},
    howpublished={RFC 3168},
    publisher={RFC Editor},
    year={2001},
    month={Sep}}

@inproceedings{Arun18,
    author={Arun, Venkat and Balakrishnan, Hari},
    title={{Copa: Practical Delay-Based Congestion Control for the Internet}},
    year={2018},
    booktitle={Proceedings of the Applied Networking Research Workshop (ANRW)},
    pages={19},
    location={Montreal, QC, Canada}}

@inproceedings{Hassani25a,
    author={Hassani, Hamid and Heemstra de Groot, Sonia M. and Niemegeers, Ignas G. and Joshi, Kishor C. and Exarchakos, George},
    title={{RAPID: Proactive Rate Control for Enabling High Data Rate and Low Latency in 5G Edge Networks}}, 
    booktitle={Proceedings of the IEEE Conference on Computer Communications (INFOCOM) Workshops}, 
    year={2025},
    pages={1-6}}

@inproceedings{Hassani25b,
    author={Hassani, Hamid and Heemstra de Groot, Sonia M. and Niemegeers, Ignas G. and Joshi, Kishor C. and Exarchakos, George},
    booktitle={Proceedings of the IEEE International Conference on Communications (ICC)}, 
    title={{One-Way Delay Model for 5G Low Latency, High Data Rate Communications}}, 
    year={2025},
    pages={5859-5865},
    location={Montreal, QC, Canada}}

@inproceedings{Larrea26,
    title={{BISCAY: Practical Radio KPI Driven Congestion Control for Mobile Networks}},
    author={Larrea, Jon and Shreedhar, Tanya and Niemi, Atte and Sefiane, Adel and Marina, Mahesh K.},
    booktitle={Proceedings of New Ideas in Networked Systems (NINeS)},
    year={2026},
    pages={15:1-15:32},
    location={Virtual Event}}

@article{scikit-learn,
    title={{Scikit-learn: Machine Learning in Python}},
    author={Pedregosa, F. and Varoquaux, G. and Gramfort, A. and Michel, V. and Thirion, B. and Grisel, O. and Blondel, M. and Prettenhofer, P. and Weiss, R. and Dubourg, V. and Vanderplas, J. and Passos, A. and Cournapeau, D. and Brucher, M. and Perrot, M. and Duchesnay, E.},
    journal={Journal of Machine Learning Research},
    volume={12},
    pages={2825-2830},
    year={2011}}

@inproceedings{Goyal20,
    author={Goyal, Prateesh and Agarwal, Anup and Netravali, Ravi and Alizadeh, Mohammad and Balakrishnan, Hari},
    booktitle={Proceedings of the USENIX Symposium on Networked Systems Design and Implementation (NSDI)},
    title={{ABC: A Simple Explicit Congestion Controller for Wireless Networks}},
    year={2020},
    pages={353-372},
    location={Santa Clara, CA, USA}}

@inproceedings{Karnam26,
    author={Karnam, Anudeep and Hassani, Hamid and Joshi, Kishor C. and Heemstra de Groot, Sonia M. and Niemegeers, Ignas G. and Exarchakos, Georgios},
    title={{PINOCCHIO: Queueing-Onset Prediction in Mobile Edge Networks Using Reservoir Computing}},
    year={2026},
    booktitle={Proceedings of the Applied Networking Research Workshop (ANRW)},
    pages={24-26},
    location={Vienna, Austria}}
\end{document}